\documentclass[final]{IEEEtran}
\usepackage[update,prepend]{epstopdf}

\usepackage{graphics}
\usepackage{multirow}
\usepackage{tikz}
\usepackage{bbm} 
\usepackage{pdfpages}
\usepackage{multirow}
\usepackage{subfig}
\usepackage{comment}
\usepackage{makecell}

\usepackage{setspace}	
\usepackage{graphicx}
\usepackage{algorithm,algorithmic}
\usepackage{multicol}

\usepackage[justification=centering]{caption}
\usepackage{textcomp}
\usepackage{psfrag}
\usepackage{arydshln}
\usepackage{url}
\usepackage{soul}
\usepackage{graphicx,color}
\usepackage[nolist]{acronym}
\usepackage{array}

\usepackage{mathtools,lipsum}
\usepackage{cuted}
\usepackage{amsmath}
\usepackage{graphicx}
\usepackage{threeparttable}

\def\nb0{{\mathbf{0}}}
\def\nb1{{\mathbf{1}}}

\begin{document}
\graphicspath{{./Figures/}}
	\begin{acronym}

\acro{5G-NR}{5G New Radio}
\acro{3GPP}{3rd Generation Partnership Project}
\acro{ABS}{aerial base station}
\acro{AC}{address coding}
\acro{ACF}{autocorrelation function}
\acro{ACR}{autocorrelation receiver}
\acro{ADC}{analog-to-digital converter}
\acrodef{aic}[AIC]{Analog-to-Information Converter}     
\acro{AIC}[AIC]{Akaike information criterion}
\acro{aric}[ARIC]{asymmetric restricted isometry constant}
\acro{arip}[ARIP]{asymmetric restricted isometry property}

\acro{ARQ}{Automatic Repeat Request}
\acro{AUB}{asymptotic union bound}
\acrodef{awgn}[AWGN]{Additive White Gaussian Noise}     
\acro{AWGN}{additive white Gaussian noise}

\acro{APSK}[PSK]{asymmetric PSK} 

\acro{waric}[AWRICs]{asymmetric weak restricted isometry constants}
\acro{warip}[AWRIP]{asymmetric weak restricted isometry property}
\acro{BCH}{Bose, Chaudhuri, and Hocquenghem}        
\acro{BCHC}[BCHSC]{BCH based source coding}
\acro{BEP}{bit error probability}
\acro{BFC}{block fading channel}
\acro{BG}[BG]{Bernoulli-Gaussian}
\acro{BGG}{Bernoulli-Generalized Gaussian}
\acro{BPAM}{binary pulse amplitude modulation}
\acro{BPDN}{Basis Pursuit Denoising}
\acro{BPPM}{binary pulse position modulation}
\acro{BPSK}{Binary Phase Shift Keying}
\acro{BPZF}{bandpass zonal filter}
\acro{BSC}{binary symmetric channels}              
\acro{BU}[BU]{Bernoulli-uniform}
\acro{BER}{bit error rate}
\acro{BS}{base station}
\acro{BW}{BandWidth}
\acro{BLLL}{ binary log-linear learning }

\acro{CP}{Cyclic Prefix}
\acrodef{cdf}[CDF]{cumulative distribution function}   
\acro{CDF}{Cumulative Distribution Function}
\acrodef{c.d.f.}[CDF]{cumulative distribution function}
\acro{CCDF}{complementary cumulative distribution function}
\acrodef{ccdf}[CCDF]{complementary CDF}               
\acrodef{c.c.d.f.}[CCDF]{complementary cumulative distribution function}
\acro{CD}{cooperative diversity}

\acro{CDMA}{Code Division Multiple Access}
\acro{ch.f.}{characteristic function}
\acro{CIR}{channel impulse response}
\acro{cosamp}[CoSaMP]{compressive sampling matching pursuit}
\acro{CR}{cognitive radio}
\acro{cs}[CS]{compressed sensing}                   
\acrodef{cscapital}[CS]{Compressed sensing} 
\acrodef{CS}[CS]{compressed sensing}
\acro{CSI}{channel state information}
\acro{CCSDS}{consultative committee for space data systems}
\acro{CC}{convolutional coding}
\acro{Covid19}[COVID-19]{Coronavirus disease}

\acro{DAA}{detect and avoid}
\acro{DAB}{digital audio broadcasting}
\acro{DCT}{discrete cosine transform}
\acro{dft}[DFT]{discrete Fourier transform}
\acro{DR}{distortion-rate}
\acro{DS}{direct sequence}
\acro{DS-SS}{direct-sequence spread-spectrum}
\acro{DTR}{differential transmitted-reference}
\acro{DVB-H}{digital video broadcasting\,--\,handheld}
\acro{DVB-T}{digital video broadcasting\,--\,terrestrial}
\acro{DL}{DownLink}
\acro{DSSS}{Direct Sequence Spread Spectrum}
\acro{DFT-s-OFDM}{Discrete Fourier Transform-spread-Orthogonal Frequency Division Multiplexing}
\acro{DAS}{Distributed Antenna System}
\acro{DNA}{DeoxyriboNucleic Acid}

\acro{EC}{European Commission}
\acro{EED}[EED]{exact eigenvalues distribution}
\acro{EIRP}{Equivalent Isotropically Radiated Power}
\acro{ELP}{equivalent low-pass}
\acro{eMBB}{Enhanced Mobile Broadband}
\acro{EMF}{ElectroMagnetic Field}
\acro{EU}{European union}
\acro{EI}{Exposure Index}
\acro{eICIC}{enhanced Inter-Cell Interference Coordination}

\acro{FC}[FC]{fusion center}
\acro{FCC}{Federal Communications Commission}
\acro{FEC}{forward error correction}
\acro{FFT}{fast Fourier transform}
\acro{FH}{frequency-hopping}
\acro{FH-SS}{frequency-hopping spread-spectrum}
\acrodef{FS}{Frame synchronization}
\acro{FSsmall}[FS]{frame synchronization}  
\acro{FDMA}{Frequency Division Multiple Access}

\acro{GA}{Gaussian approximation}
\acro{GF}{Galois field }
\acro{GG}{Generalized-Gaussian}
\acro{GIC}[GIC]{generalized information criterion}
\acro{GLRT}{generalized likelihood ratio test}
\acro{GPS}{Global Positioning System}
\acro{GMSK}{Gaussian Minimum Shift Keying}
\acro{GSMA}{Global System for Mobile communications Association}
\acro{GS}{ground station}
\acro{GMG}{ Grid-connected MicroGeneration}

\acro{HAP}{high altitude platform}
\acro{HetNet}{Heterogeneous network}

\acro{IDR}{information distortion-rate}
\acro{IFFT}{inverse fast Fourier transform}
\acro{iht}[IHT]{iterative hard thresholding}
\acro{i.i.d.}{independent, identically distributed}
\acro{IoT}{Internet of Things}                      
\acro{IR}{impulse radio}
\acro{lric}[LRIC]{lower restricted isometry constant}
\acro{lrict}[LRICt]{lower restricted isometry constant threshold}
\acro{ISI}{intersymbol interference}
\acro{ITU}{International Telecommunication Union}
\acro{ICNIRP}{International Commission on Non-Ionizing Radiation Protection}
\acro{IEEE}{Institute of Electrical and Electronics Engineers}
\acro{ICES}{IEEE international committee on electromagnetic safety}
\acro{IEC}{International Electrotechnical Commission}
\acro{IARC}{International Agency on Research on Cancer}
\acro{IS-95}{Interim Standard 95}

\acro{KPI}{Key Performance Indicator}

\acro{LEO}{low earth orbit}
\acro{LF}{likelihood function}
\acro{LLF}{log-likelihood function}
\acro{LLR}{log-likelihood ratio}
\acro{LLRT}{log-likelihood ratio test}
\acro{LoS}{Line-of-Sight}
\acro{LRT}{likelihood ratio test}
\acro{wlric}[LWRIC]{lower weak restricted isometry constant}
\acro{wlrict}[LWRICt]{LWRIC threshold}
\acro{LPWAN}{Low Power Wide Area Network}
\acro{LoRaWAN}{Low power long Range Wide Area Network}
\acro{NLoS}{Non-Line-of-Sight}
\acro{LiFi}[Li-Fi]{light-fidelity}
 \acro{LED}{light emitting diode}
 \acro{LABS}{LoS transmission with each ABS}
 \acro{NLABS}{NLoS transmission with each ABS}

\acro{MB}{multiband}
\acro{MC}{macro cell}
\acro{MDS}{mixed distributed source}
\acro{MF}{matched filter}
\acro{m.g.f.}{moment generating function}
\acro{MI}{mutual information}
\acro{MIMO}{Multiple-Input Multiple-Output}
\acro{MISO}{multiple-input single-output}
\acrodef{maxs}[MJSO]{maximum joint support cardinality}                       
\acro{ML}[ML]{maximum likelihood}
\acro{MMSE}{minimum mean-square error}
\acro{MMV}{multiple measurement vectors}
\acrodef{MOS}{model order selection}
\acro{M-PSK}[${M}$-PSK]{$M$-ary phase shift keying}                       
\acro{M-APSK}[${M}$-PSK]{$M$-ary asymmetric PSK} 
\acro{MP}{ multi-period}
\acro{MINLP}{mixed integer non-linear programming}

\acro{M-QAM}[$M$-QAM]{$M$-ary quadrature amplitude modulation}
\acro{MRC}{maximal ratio combiner}                  
\acro{maxs}[MSO]{maximum sparsity order}                                      
\acro{M2M}{Machine-to-Machine}                                                
\acro{MUI}{multi-user interference}
\acro{mMTC}{massive Machine Type Communications}      
\acro{mm-Wave}{millimeter-wave}
\acro{MP}{mobile phone}
\acro{MPE}{maximum permissible exposure}
\acro{MAC}{media access control}
\acro{NB}{narrowband}
\acro{NBI}{narrowband interference}
\acro{NLA}{nonlinear sparse approximation}
\acro{NLOS}{Non-Line of Sight}
\acro{NTIA}{National Telecommunications and Information Administration}
\acro{NTP}{National Toxicology Program}
\acro{NHS}{National Health Service}

\acro{LOS}{Line of Sight}

\acro{OC}{optimum combining}                             
\acro{OC}{optimum combining}
\acro{ODE}{operational distortion-energy}
\acro{ODR}{operational distortion-rate}
\acro{OFDM}{Orthogonal Frequency-Division Multiplexing}
\acro{omp}[OMP]{orthogonal matching pursuit}
\acro{OSMP}[OSMP]{orthogonal subspace matching pursuit}
\acro{OQAM}{offset quadrature amplitude modulation}
\acro{OQPSK}{offset QPSK}
\acro{OFDMA}{Orthogonal Frequency-division Multiple Access}
\acro{OPEX}{Operating Expenditures}
\acro{OQPSK/PM}{OQPSK with phase modulation}

\acro{PAM}{pulse amplitude modulation}
\acro{PAR}{peak-to-average ratio}
\acrodef{pdf}[PDF]{probability density function}                      
\acro{PDF}{probability density function}
\acrodef{p.d.f.}[PDF]{probability distribution function}
\acro{PDP}{power dispersion profile}
\acro{PMF}{probability mass function}                             
\acrodef{p.m.f.}[PMF]{probability mass function}
\acro{PN}{pseudo-noise}
\acro{PPM}{pulse position modulation}
\acro{PRake}{Partial Rake}
\acro{PSD}{power spectral density}
\acro{PSEP}{pairwise synchronization error probability}
\acro{PSK}{phase shift keying}
\acro{PD}{power density}
\acro{8-PSK}[$8$-PSK]{$8$-phase shift keying}
\acro{PPP}{Poisson point process}
\acro{PCP}{Poisson cluster process}
 
\acro{FSK}{Frequency Shift Keying}

\acro{QAM}{Quadrature Amplitude Modulation}
\acro{QPSK}{Quadrature Phase Shift Keying}
\acro{OQPSK/PM}{OQPSK with phase modulator }

\acro{RD}[RD]{raw data}
\acro{RDL}{"random data limit"}
\acro{ric}[RIC]{restricted isometry constant}
\acro{rict}[RICt]{restricted isometry constant threshold}
\acro{rip}[RIP]{restricted isometry property}
\acro{ROC}{receiver operating characteristic}
\acro{rq}[RQ]{Raleigh quotient}
\acro{RS}[RS]{Reed-Solomon}
\acro{RSC}[RSSC]{RS based source coding}
\acro{r.v.}{random variable}                               
\acro{R.V.}{random vector}
\acro{RMS}{root mean square}
\acro{RFR}{radiofrequency radiation}
\acro{RIS}{Reconfigurable Intelligent Surface}
\acro{RNA}{RiboNucleic Acid}
\acro{RRM}{Radio Resource Management}
\acro{RUE}{reference user equipments}
\acro{RAT}{radio access technology}
\acro{RB}{resource block}

\acro{SA}[SA-Music]{subspace-augmented MUSIC with OSMP}
\acro{SC}{small cell}
\acro{SCBSES}[SCBSES]{Source Compression Based Syndrome Encoding Scheme}
\acro{SCM}{sample covariance matrix}
\acro{SEP}{symbol error probability}
\acro{SG}[SG]{sparse-land Gaussian model}
\acro{SIMO}{single-input multiple-output}
\acro{SINR}{signal-to-interference plus noise ratio}
\acro{SIR}{signal-to-interference ratio}
\acro{SISO}{Single-Input Single-Output}
\acro{SMV}{single measurement vector}
\acro{SNR}[\textrm{SNR}]{signal-to-noise ratio} 
\acro{sp}[SP]{subspace pursuit}
\acro{SS}{spread spectrum}
\acro{SW}{sync word}
\acro{SAR}{specific absorption rate}
\acro{SSB}{synchronization signal block}
\acro{SR}{shrink and realign}

\acro{tUAV}{tethered Unmanned Aerial Vehicle}
\acro{TBS}{terrestrial base station}

\acro{uUAV}{untethered Unmanned Aerial Vehicle}
\acro{PDF}{probability density functions}

\acro{PL}{path-loss}

\acro{TH}{time-hopping}
\acro{ToA}{time-of-arrival}
\acro{TR}{transmitted-reference}
\acro{TW}{Tracy-Widom}
\acro{TWDT}{TW Distribution Tail}
\acro{TCM}{trellis coded modulation}
\acro{TDD}{Time-Division Duplexing}
\acro{TDMA}{Time Division Multiple Access}
\acro{Tx}{average transmit}

\acro{UAV}{Unmanned Aerial Vehicle}
\acro{uric}[URIC]{upper restricted isometry constant}
\acro{urict}[URICt]{upper restricted isometry constant threshold}
\acro{UWB}{ultrawide band}
\acro{UWBcap}[UWB]{Ultrawide band}   
\acro{URLLC}{Ultra Reliable Low Latency Communications}
         
\acro{wuric}[UWRIC]{upper weak restricted isometry constant}
\acro{wurict}[UWRICt]{UWRIC threshold}                
\acro{UE}{User Equipment}
\acro{UL}{UpLink}

\acro{WiM}[WiM]{weigh-in-motion}
\acro{WLAN}{wireless local area network}
\acro{wm}[WM]{Wishart matrix}                               
\acroplural{wm}[WM]{Wishart matrices}
\acro{WMAN}{wireless metropolitan area network}
\acro{WPAN}{wireless personal area network}
\acro{wric}[WRIC]{weak restricted isometry constant}
\acro{wrict}[WRICt]{weak restricted isometry constant thresholds}
\acro{wrip}[WRIP]{weak restricted isometry property}
\acro{WSN}{wireless sensor network}                        
\acro{WSS}{Wide-Sense Stationary}
\acro{WHO}{World Health Organization}
\acro{Wi-Fi}{Wireless Fidelity}

\acro{sss}[SpaSoSEnc]{sparse source syndrome encoding}

\acro{VLC}{Visible Light Communication}
\acro{VPN}{Virtual Private Network} 
\acro{RF}{Radio Frequency}
\acro{FSO}{Free Space Optics}
\acro{IoST}{Internet of Space Things}

\acro{GSM}{Global System for Mobile Communications}
\acro{2G}{Second-generation cellular network}
\acro{3G}{Third-generation cellular network}
\acro{4G}{Fourth-generation cellular network}
\acro{5G}{Fifth-generation cellular network}	
\acro{gNB}{next-generation Node-B Base Station}
\acro{NR}{New Radio}
\acro{UMTS}{Universal Mobile Telecommunications Service}
\acro{LTE}{Long Term Evolution}

\acro{QoS}{Quality of Service}
\end{acronym}
	
\newcommand{\SAR} {\mathrm{SAR}}
\newcommand{\WBSAR} {\mathrm{SAR}_{\mathsf{WB}}}
\newcommand{\gSAR} {\mathrm{SAR}_{10\si{\gram}}}
\newcommand{\Sab} {S_{\mathsf{ab}}}
\newcommand{\Eavg} {E_{\mathsf{avg}}}
\newcommand{\ft}{f_{\textsf{th}}}
\newcommand{\alphatf}{\alpha_{24}}

\title{
Network-Level Analysis of Integrated Sensing and Communication Using Stochastic Geometry
}
\author{
 
Ruibo Wang, Baha Eddine Youcef Belmekki, {\em Member, IEEE}, \\ Xue Zhang, and Mohamed-Slim Alouini, {\em Fellow, IEEE}
\thanks{The authors are with King Abdullah University of Science and Technology (KAUST), CEMSE division, Thuwal 23955-6900, Saudi Arabia (e-mail: ruibo.wang@kaust.edu.sa; 
bahaeddine.belmekki@kaust.edu.sa; xue.zhang@kaust.edu.sa;
slim.alouini@kaust.edu.sa).}
\vspace{-6mm}
}
\maketitle
\thispagestyle{empty}
\begin{abstract}
To meet the demands of densely deploying communication and sensing devices in the next generation of wireless networks, integrated sensing and communication (ISAC) technology is employed to alleviate spectrum scarcity, while stochastic geometry (SG) serves as a tool for low-complexity performance evaluation. To assess network-level performance, there is a natural interaction between ISAC technology and the SG method. From ISAC network perspective, we illustrate how to leverage SG analytical framework to evaluate ISAC network performance by introducing point process distributions and stochastic fading channel models. From SG framework perspective, we summarize the unique performance metrics and research objectives of ISAC networks, thereby extending the scope of SG research in the field of wireless communications. Additionally, considering the limited discussion in the existing SG-based ISAC works in terms of distribution and channel modeling, a case study is designed to exploit topology and channel fading awareness to provide relevant network insights. 
\end{abstract}

\section{Introduction}\label{section1}
The arrival of the next generation of wireless networks and the explosive growth in the number of wireless communication devices have massively increased the demand for wireless spectrum \cite{huang2023system,lou2023haps}. { Emerging networks are anticipated to surpass the constraints of communication functionalities, extending their capabilities to include sensing functionalities for measuring and imaging the surrounding environment \cite{liu2022integrated}. These applications span various domains, including smart cities, intelligent transportation systems, and the internet of things (IoT).} 

\par
However, incorporating sensing capabilities into networks requires additional spectrum support, energy consumption, and hardware infrastructure. Therefore, the limited resources of next-generation networks and the high demand for communication and sensing have given rise to the concept of integrated sensing and communication (ISAC) networks~\cite{salem2022rethinking}. ISAC networks enable the sharing of spectrum and hardware platforms between communication and sensing, thereby significantly enhancing spectrum, energy, and hardware efficiency~\cite{ma2022performance}. More importantly, ISAC can also facilitate new integrated modes from a collaborative design perspective, enabling mutual benefits between communication and sensing.

Traditional IoT networks have additional sensing capabilities with communication functions, which are changing from separation to integration. From technical perspective, the sensing and communication devices can share network resources with a unified framework. From commercial perspective, this integration promotes the emergence of new IoT services and applications. Consequently, these ISAC-enabled IoT networks will play a key role in next-generation networks. Due to the need for analyzing sensing and communication performance and balancing the allocation of network resources, performance analysis of ISAC is crucial. Also, in terms of economy of scale, particularly in the automotive industry, combining sensing and communication functions into a single hardware significantly reduces the costs associated with manufacturing and installing multiple separate systems in vehicles.

For large-scale IoT networks, studying the overall performance is often more relevant and insightful compared to investigating the behavior of individual IoT devices. In addition, the dense deployment of a massive number of IoT devices makes interference a critical factor that affects ISAC network performance~\cite{ali2022integrated}. Stochastic geometry (SG) is one of the most suitable mathematical tools for large-scale network modeling and interference analysis \cite{wang2022ultra}. It requires only a few parameters, such as densities of communication devices and sensing devices, to construct a stochastic network topology and provide analytical results for performance. In the analysis of large-scale network performance, compared to simulation that requires a large amount of data, SG is a low-cost, accurate, and computationally inexpensive alternative \cite{wang2022stochastic}.
{ 
The contributions of this magazine paper are as follows:
\begin{itemize}
    \item We have summarized existing research from a special perspective of addressing two types of readers: one group comprises readers from the ISAC domain who seek to understand how to utilize the SG analytical framework for performance analysis, while the other group consists of readers from the SG domain interested in learning how to extend traditional wireless networks to ISAC networks.
    \item Regarding the unique challenges faced by ISAC-SG research and the limitations of the existing studies, we have put forward our insights and identified three potential directions for future research about modeling, metric, and scenario.
    \item We proposed a case study of a sensing-assisted communication system with a joint base stations (BSs) and unmanned aerial vehicles (UAVs) network. This case study addresses the shortcomings in existing research regarding modeling accuracy. It comprehensively analyzes four key communication and sensing metrics, along with a comprehensive performance evaluation parameter proposed in this magazine paper.
\end{itemize}
}

\begin{figure*}[t]
\centering
\vspace{-0.2cm}
\includegraphics[width = \textwidth]{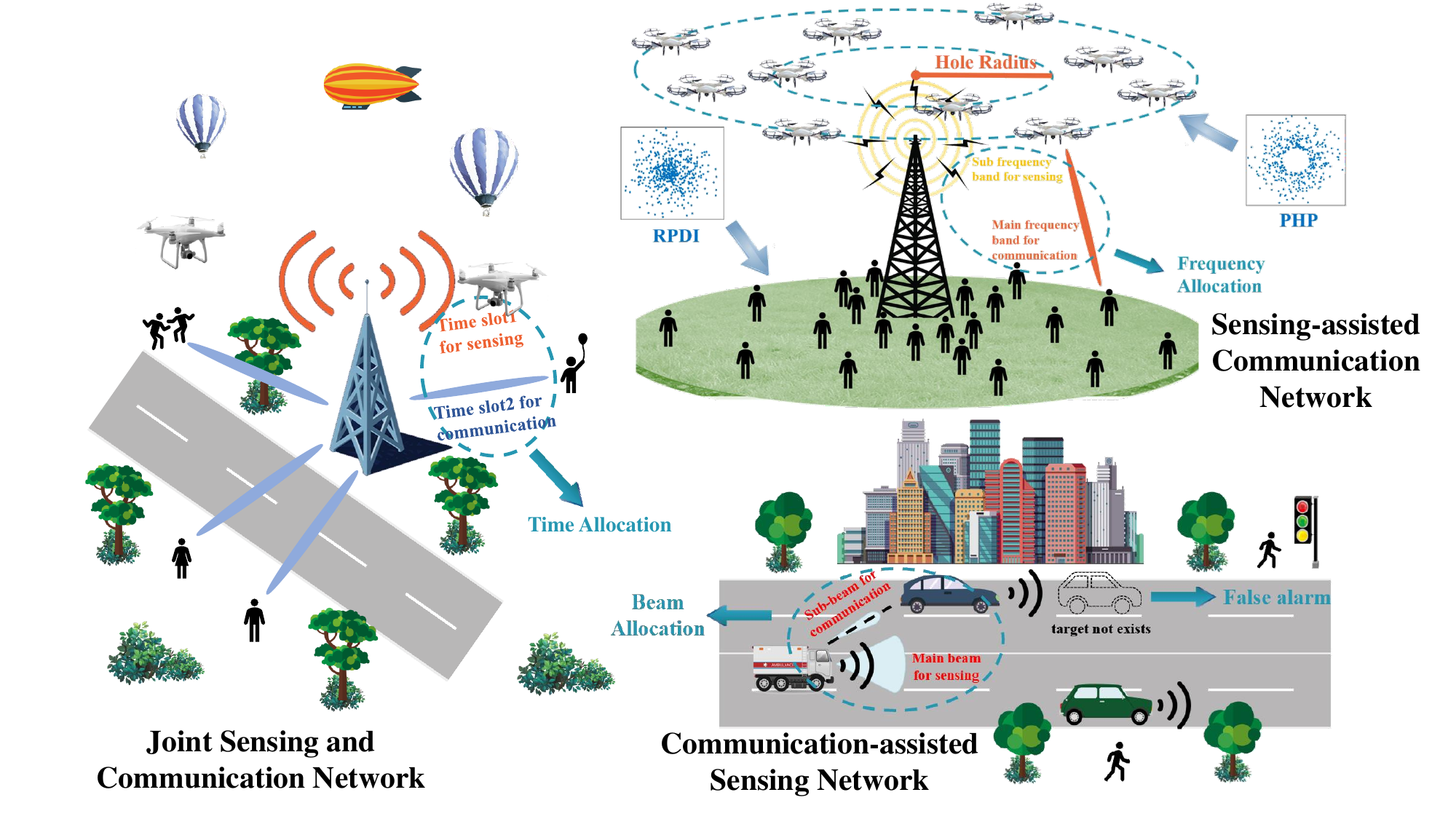}
\caption{ISAC network structure diagram.}
\vspace{-0.2cm}
\label{figure1}
\end{figure*}

\section{On the Use of SG Modeling into ISAC Paradigm} \label{section2}
This section addresses researchers in the ISAC field. We will summarize how to apply distribution and channel models within the SG framework for ISAC studies. Overall, to facilitate analysis, existing studies provided in Table~\ref{table1} have employed widely-used but over-simplified distribution and channel models. Therefore, we provide our opinions in three typical scenarios.


\subsection{Spatial Distribution}
The existing literature primarily models transmitter and receiver with homogeneous Poisson point processes (PPPs). { In PPP, the points' density is homogeneous, and the position of a point in PPP is independent of other points. The assumptions of homogeneity and independence can bring convenience for analytical derivations \cite{lou2023coverage}. General communicating networks often model either the transmitters or the receivers as a single PPP. While in ISAC networks, the spatial distribution is more complex.

\par
Due to the increase in the type of devices, the network may include communication transmitters, detection transmitters, users receiving communication services, and sensed targets \cite{salem2022rethinking}. In this case, the spatial distribution of devices in the ISAC network may be modeled as up to four independent PPPs. Complex spatial distribution models may lead to a more intricate analytical process. For instance, an increase in the density of sensing devices may result in more resources being allocated to the sensing network, subsequently affecting the performance of the communication network.
}

\begin{table*}
\centering
\vspace{-0.2cm}
\begin{threeparttable}[b]
\small
\caption{Classification and summary of existing studies.}
\label{table1}
\renewcommand{\arraystretch}{1.4}
\begin{tabular}{|c|c|c|c|c|c|c|} 
\hline 
Ref. & Type & System & Model & \makecell[c]{Small-scale \\ fading} & Metric & Objective \\ 
\cline{1-7}
\cite{salem2022rethinking} & Cellular & JSC & PPP & Rayleigh & \makecell[c]{(J) Potential spectral efficiency \\ (PSE), energy efficiency} & Optimization \\
\cline{1-7}
\cite{ma2022performance} & Vehicular & JSC & PPP & N/A & \makecell[c]{(S) Detection probability; \\ (C) Coverage probability} & \makecell[c]{Performance  evaluation} \\
\cline{1-7}
\cite{ali2022integrated} & Cellular & JSC & PPP & Rayleigh & \makecell[c]{(S) SINR; (C) Throughput} & \makecell[c]{Resource allocation} \\
\cline{1-7}
\cite{fang2019performance} & N/A & C-AS & PPP & Rayleigh & (S) Detection range & \makecell[c]{Resource allocation} \\ 
\cline{1-7}
\cite{ghozlani2021stochastic} & Vehicular & C-AS & PPP & N/A & (S) Detection range & \makecell[c]{Performance evaluation} \\  
\cline{1-7}
\cite{ram2022optimization} & Cellular & S-AC & PPP & N/A & (C) Throughput & 
Optimization \\ 
\cline{1-7}
\cite{moulin2022joint} & Vehicular & JSC & PPP & Rayleigh & \makecell[c]{(S) Detection probability; \\ (C) Throughput} & \makecell[c]{Performance evaluation} \\
\cline{1-7}
\cite{olson2022coverage} & Cellular & JSC & PPP & Rayleigh & \makecell[c]{ (J) Extended coverage probability, \\ extended throughput} & \makecell[c]{Performance evaluation} \\
\cline{1-7}
\cite{chen2022isac} & N/A & S-AC & PPP & N/A & \makecell[c]{(C) Coverage probability, \\ throughput} & \makecell[c]{Performance evaluation} \\
\cline{1-7}
\cite{skouroumounis2021fd} & Cellular & C-AS & GPP \footnote{} & Nakagami-$m$ & \makecell[c]{(S) Detection probability, \\ false alarm probability} & \makecell[c]{Performance evaluation} \\
\cline{1-7}
\makecell[c]{\cite{ren2018performance} \\ \cite{ren2020performance}} & N/A & JSC & PPP & N/A & \makecell[c]{(S) Detection probability, false \\ alarm probability;  (C) Throughput} & \makecell[c]{Resource allocation} \\  
\cline{1-7}
\cite{maeng2022analysis} & UAV & JSC & PPP & Rician & \makecell[c]{(S) Detection probability; \\ (C) Throughput} & \makecell[c]{Resource allocation} \\
\hline 
\end{tabular}
\begin{footnotesize}
\begin{tablenotes}
    \item [1] GPP refers to the Ginibre point process.
\end{tablenotes}
\end{footnotesize}
\end{threeparttable}
\end{table*}

\subsection{Channel Model}
The current works in Table~\ref{table1} follow generic but simplified channel models. The received signal power is modeled as the product of shadowing, antenna gain, path loss, and small-scale fading. {  Currently, relevant literature models antenna gain as omnidirectional gain \cite{ma2022performance}. However, the antenna gain in sensing networks is typically characterized as directional to mitigate interference. Specifically, targets closer to the detection direction will experience larger beam gains and cause stronger interference. Considering that modeling directional antenna gain plays a significant role in accurately evaluating the performance of interference-dominant networks, the transmitter is equipped with a Gaussian beam directional model in the case study of this article. The heterogeneity modeling of antenna gain and interference power renders traditional analytical methods in communication networks, utilizing Laplace transforms for interference, no longer applicable. As a unique challenge to ISAC networks, integrating directional beams into the channel model will serve as a meaningful future research direction.}

\par
{  Another shortcoming in existing literature lie in addressing complex terrain environments, since the currently studied ISAC-SG application scenarios are typical instances of obstruction-sensitive environments. Authors in \cite{skouroumounis2021fd}  employ Nakagami-$m$ distribution in modeling small-scale fading to describe multi-path effects in complex terrain situations. In addition, the authors in \cite{ram2022optimization} emphasize the necessity of considering obstacles in the future outlook. This proposal undoubtedly poses challenges for modeling and analysis, yet it is indispensable for the performance analysis of ISAC networks. Therefore, we take building blockage into consideration in the communication link modeling of this case study.}

\subsection{Typical Application Scenarios}
Considering that existing works have employed overly simplified distribution and channel models, they have fallen short of accurately capturing the topology and fading characteristics of real-world ISAC networks. Therefore, drawing inspiration from modeling methodologies employed in general wireless networks, we offer suggestions for refining the modeling approaches in three typical scenarios, while also outlining the motivation behind the case study. 

\subsubsection{Cellular Networks}
In cellular networks, communication or sensing signal transmitters are not extensively deployed in close proximity to avoid resource wastage and mutual interference. Therefore, they are suitable for employing exclusion processes with minimum distance constraints, such as the Matérn hardcore point process (MHCPP). Conversely, to fulfill the requirements of high-quality communication and precise detection, users and targets tend to be distributed around signal transmitters, making them suitable for modeling by Poisson cluster processes (PCPs).

\subsubsection{Vehicular Networks}
{  As a crucial auxiliary for autonomous vehicles, ISAC technology can provide environmental information and is expected to extend the vehicle's perception range beyond its field of view \cite{liu2022integrated}.} For homogeneity, vehicles are distributed along roads, therefore more suitable to be modeled as a locally homogeneous PPP within strip-shaped regions, referred to as the Cox point process. Vehicles are typically modeled in complex urban environments, where Nakagami-$m$ fading is more suitable for small-scale fading modeling \cite{skouroumounis2021fd}.

\subsubsection{UAV Networks}
UAVs are typically employed to enhance communication or sensing capabilities in core cellular networks. As a result, UAVs are typically not deployed close to signal sources, but rather in cellular edge areas that are challenging for signal sources to cover or detect. In such cases, the Poisson hole process (PHP) is one of the most suitable models for UAV networks.  

\par
{  
Finally, schematic diagrams for cellular networks, vehicular networks, and UAV networks are respectively presented on the left, bottom right, and top right parts of Fig.~\ref{figure1}.}

\section{Extend ISAC Context Within the SG Framework}\label{section3}
This section explains how to extend the existing SG-based wireless communication network analysis to SG-based ISAC network analysis. Overall, researchers have conducted comprehensive studies on network types, performance metrics, and research objectives of ISAC networks under the SG framework. We will provide further elaboration according to the summary and classification in Table~\ref{table1}.

\subsection{Network Type}\label{sec1-2}
This subsection discusses three types of networks called communication-assisted sensing networks, sensing-assisted communication networks, and joint sensing and communication networks, which are respectively denoted as C-AS, S-AC, and JSC in Table~\ref{table1}. {  In addition, we have respectively presented schematic diagrams of C-AS, S-AC, and JSC in the bottom right, top right, and left parts of Fig.~\ref{figure1}.}

\subsubsection{Communication-Assisted Sensing Networks} 
{ The concepts of C-AS and S-AC were first proposed by the authors in \cite{liu2022integrated}.} Due to the limited sensing capabilities of individual devices, multiple sensing devices can form a C-AS network. The network expands the sensing coverage and improves sensing performance by sharing and exchanging real-time sensing messages. As shown in the bottom-right part of Fig~\ref{figure1}, in vehicular ad-hoc networks (VANETs) \cite{ma2022performance}, onboard radars in vehicles have limited detection range thus cause blind spots, which introduces the risk of missed detection for distant targets \cite{fang2019performance}. To address this issue, vehicles can share detection information, thereby expanding the cooperative detection range. Considering that VANETs are large-scale dynamic networks, SG methods are suitable for distance analysis in dynamic network topologies. By leveraging concepts under the SG framework, such as contact distance distribution, authors in \cite{ghozlani2021stochastic} derive closed-form expressions for the cooperative detection distance between vehicles, and investigate the cooperative detection performances under different network configurations.

\subsubsection{Sensing-Assisted Communication Networks}
Similar to how C-AS networks exploit communication sharing nature to obtain additional information for sensing performance improvement, S-AC networks acquire additional environmental information to enhance communication quality by utilizing sensing capabilities. For example, user terminals may experience blockage from obstacles, leading to leaving the line-of-sight range of communication devices or losing connectivity due to distance increases \cite{ram2022optimization}. The communication devices can rely on radar sensing capabilities to predict the positions of users, track their movement trajectories, and adjust communication strategies accordingly. Unlike the previous work \cite{ghozlani2021stochastic} about VANETs that favor SG capabilities in random topology analysis, the authors in \cite{ram2022optimization} emphasize the capability of SG analysis to decipher the random small-scale fading distribution in complex environments. Subsequently, they quantitatively estimate the effectiveness of enhancing communication performance through radar sensing.

\subsubsection{Joint Sensing and Communication Networks}
As the main direction integration between sensing and communication, one of the most typical features of ISAC networks is the balanced allocation of resources. A case of resource imbalance is using the main beam for sensing and the sub-beams for communication \cite{fang2019performance}. In the research of ISAC networks, scholars \cite{moulin2022joint} and \cite{olson2022coverage} have designed some metrics that are applicable to both sensing and communication performance evaluation. With the SG framework, they have derived analytical expressions for these metrics.

\subsection{Performance Metric}\label{sec2-3}
Table~\ref{table1} mentions three kinds of metrics: sensing metrics, communication metrics, and JSC metrics, which are abbreviated as (S), (C), and (J), respectively.

\subsubsection{Signal-to-Interference-plus-Noise Ratio (SINR)}
We first start by introducing SINR as an intermediary metric of the following metrics. Considering the strong correlation between SG and SINR, we will now introduce SINR-based sensing metrics, communication metrics, and ISAC metrics. It is worth noting that interference in some studies only refers to self-interference within the sensing network or communication network \cite{ali2022integrated}. Several studies about JSC networks take the impact of mutual interference between these two types of networks into account \cite{ren2018performance}. 

\par
\subsubsection{Sensing Metrics}
The two most widely studied SINR-related sensing metrics are detection probability and false alarm probability \cite{maeng2022analysis}. Detection probability is the probability that a sensing system can successfully detect a target when it actually exists. When the received signal is obscured by interference and noise, the target might go undetected, hence the detection probability is a function of SINR. Next, false alarm probability refers to the probability that the sensing system generates target detection alerts erroneously in the absence of an actual target. False alarms are typically caused by interference or noise being erroneously identified as a target signal, hence the false alarm probability is related to the power of interference and noise.

\par
\subsubsection{Communication Metrics}
The communication metrics in ISAC networks are similar to those in SG-based traditional wireless communication networks. Two typical communication metrics are coverage probability and throughput \cite{chen2022isac}. Coverage probability represents the probability that the received signal can be successfully demodulated and is mathematically defined as the probability of SINR exceeding a predefined threshold. Throughput indicates the communication system's ability to support data traffic and is mathematically defined as the achievable data transmission rate given by the Shannon-Hartley formula.

\subsubsection{Joint Metrics}
Joint metrics are metrics that can measure both sensing and communication performance. They can be extensions of existing metrics, such as the extended coverage probability and extended throughput defined in \cite{olson2022coverage}, making them applicable to the evaluation of sensing performance as well. Joint metrics can also be specifically defined for the performance evaluation of ISAC networks, such as the potential spectral efficiency (PSE) \cite{salem2022rethinking}. PSE is the number of bits successfully transmitted per unit area and per unit time, which is applicable to the performance evaluation of both communication and sensing networks. To comprehensively consider the needs of sensing and communication networks, Joint metrics may involve a combination or weighted sum of sensing and communication metrics. For example, in \cite{salem2022rethinking}, the sum of the PSE for the sensing network and the PSE for the communication network, divided by the total power consumption of communication and sensing, is defined as energy efficiency to evaluate the overall performance of the ISAC network.

\subsection{Research Objective}\label{sec2-4}

\subsubsection{Performance Evaluation}
Performance evaluation is the most common aspect of traditional SG research in the wireless communication domain. There is a correspondence between the types of systems and the types of metrics being evaluated. In the study of S-AC networks, the SG framework is primarily employed to analyze communication metrics such as coverage probability and throughput \cite{chen2022isac}. The impact of sensing parameters on communication performance is usually briefly discussed only in S-AC networks. Vice versa in the case of C-AS networks.

\subsubsection{Resource Allocation}
{  As shown in Fig.~\ref{figure1} resource allocation refers to the allocation of limited resources such as time, frequency, and beam/spatial resources to sensing and communication functions. Time allocation refers to the implementation of time-division multiplexing for sensing and communication in the same frequency band (in the left part of Fig.~\ref{figure1}), or the dynamic adjustment of resources allocated to sensing and communication over time \cite{maeng2022analysis} (in the upper right corner of Fig.~\ref{figure1}).} From the perspective of frequency, some literature allocates different spectra to sensing and communication networks to avoid mutual interference between the two networks, at the cost of reducing spectrum efficiency. { Finally, an illustrative example of beam allocation is found in communication-assisted sensing networks at the right bottom of Fig.~\ref{figure1}, where the main beam is utilized for sensing functions while sub-beams are used for communication \cite{fang2019performance}.}

\subsubsection{Optimization}
Optimization means adjusting the system parameters and channel parameters to achieve better sensing and communication performances. Firstly, the analytical expressions of performance metrics are derived using the SG analytical framework. In communication-assisted sensing networks, the analytical expression corresponding to the sensing metric is used as the objective function for optimization, with the communication metrics serving as constraints \cite{ram2022optimization}, and vice versa. The optimization parameters typically selected are those that simultaneously impact both sensing and communication performance, such as BS density \cite{salem2022rethinking}.


\section{Exploiting Topology and Channel Fading Awareness for ISAC} \label{section4}
As mentioned, simplified distribution and channel models in existing studies limit the full potential of SG in performance analysis. In this case study, we will exploit topology and channel fading awareness for ISAC and study the performance of an S-AC network by numerical results.

\subsection{System Model} 
We are considering deploying a BS at the community center with an altitude of $50$ meters to provide network coverage for local ground residents. Several UAVs are deployed to extend the coverage range of the BS and enhance coverage for residents at the edge of the community. Conversely, BS controls UAVs to achieve fine-grained networking and collision avoidance.

\subsubsection{Resident Distribution}
The ground residents within a circular area with a radius of $1.5$\,km around the community center/BS follow a resident population density-inspired (RPDI) model \cite{wang2023resident}. { As shown in the upper right corner of Fig.~\ref{figure1}, the RPDI model is a type of non-homogeneous PPP, whose density is proportional to $\exp(-\beta r)$. $\beta$ is known as the non-homogeneity, and $r$ is the distance to the community center.}

\subsubsection{UAV Distribution}
{  As shown in the upper right corner of Fig.~\ref{figure1}, UAVs are distributed at a fixed altitude within the same circular area as the residents.} We assume UAVs follow a similar distribution according to the RPDI model since the authors in \cite{wang2023resident} have demonstrated that this approach effectively alleviates the pressure on communication capacity. The only difference between the two distributions is that UAVs will not be deployed near the BS, thereby preventing interference with the BS. { Specifically, UAVs will not be deployed within a circular area centered around the projection of the BS onto the UAV deployment plane, and an example is given at the upper right corner of Fig.~\ref{figure1}.} 
The radius of this circular area is called the hole radius. Note that the hole radius numerically corresponds to the projection distance rather than the Euclidean distance. Overall, the UAV distribution follows a part of a non-homogeneous Poisson hole process (PHP) with a single hole.

\subsubsection{Channel Channel}
Firstly, the BS-resident and UAV-resident downlink communication links both follow the channel models experienced with Nakagami-$m$ fading. {  According to \cite{wang2023resident}, UAV-resident communication link takes building blockage into consideration. The probability of the UAV-resident link being blocked is modeled as a statistical model related to the UAV's elevation angle. Once identified as blocked, the channel's path-loss exponent will increase from $2$ to $3$, and the received signal will experience an additional $20$~dB attenuation. The values of the communication links' channel parameters refer to \cite{wang2023resident}. It is important to note that the signal is transmitted at $2$~GHz. Therefore, the BS is equipped with a radio frequency detector.} Secondly, the BS-UAV sensing channel considers Rayleigh small-scale fading \cite{moulin2022joint}. In the sensing channel, a BS transmits directional signals with a Gaussian beam model toward UAVs and receives the reflected signals.

{ 
\subsubsection{Association Strategy}
Residents follow the strongest average received power association strategy. When the capacity of a certain UAV reaches 20~Mbps, the UAV will disassociate from the farthest residents. At this point, the disassociated residents need to connect to a device with the strongest average received power among the remaining capacity-unsaturated UAVs and the BS. We simulate using the Monte Carlo method \cite{wang2023resident}. In each simulation round, we generate a PHP and an RPDI. In other words, the positions of UAVs and residents are regenerated in each simulation round, and the associations between UAVs and residents are also regenerated. This approach allows us to simulate the hand-off process of UAVs and the movement of UAVs and residents. For residents, apart from the expected signal power provided by the associated UAV or BS, the received power from other devices is considered as interference. For the UAV-BS link, apart from the signal from the UAV aligned with the BS, the received power from other UAVs is considered as interference. Due to the directional beam model, UAVs deviating significantly from the alignment direction receive minimal antenna gain, resulting in limited interference. Based on the expected signal and interference power, we can compute SINR-related metrics. From the perspective of the SG analytical framework, given the association strategy mentioned above, we can derive the CDF or analytical expressions for SINR-related metrics. These expressions can be formulated as functions of system parameters such as hole radius and UAV height, achieving a low-complexity mapping from system parameters to metrics. 
}

\subsection{Communication Metric}\label{section4-2}
In the subsequent numerical results, we optimize the communication and sensing performances by adjusting the parameters of UAV distribution. As mentioned, the parameters include the UAV height, hole radius, and non-homogeneity. 

\par
\begin{figure}[t]
\centering
\includegraphics[width=0.9\linewidth]{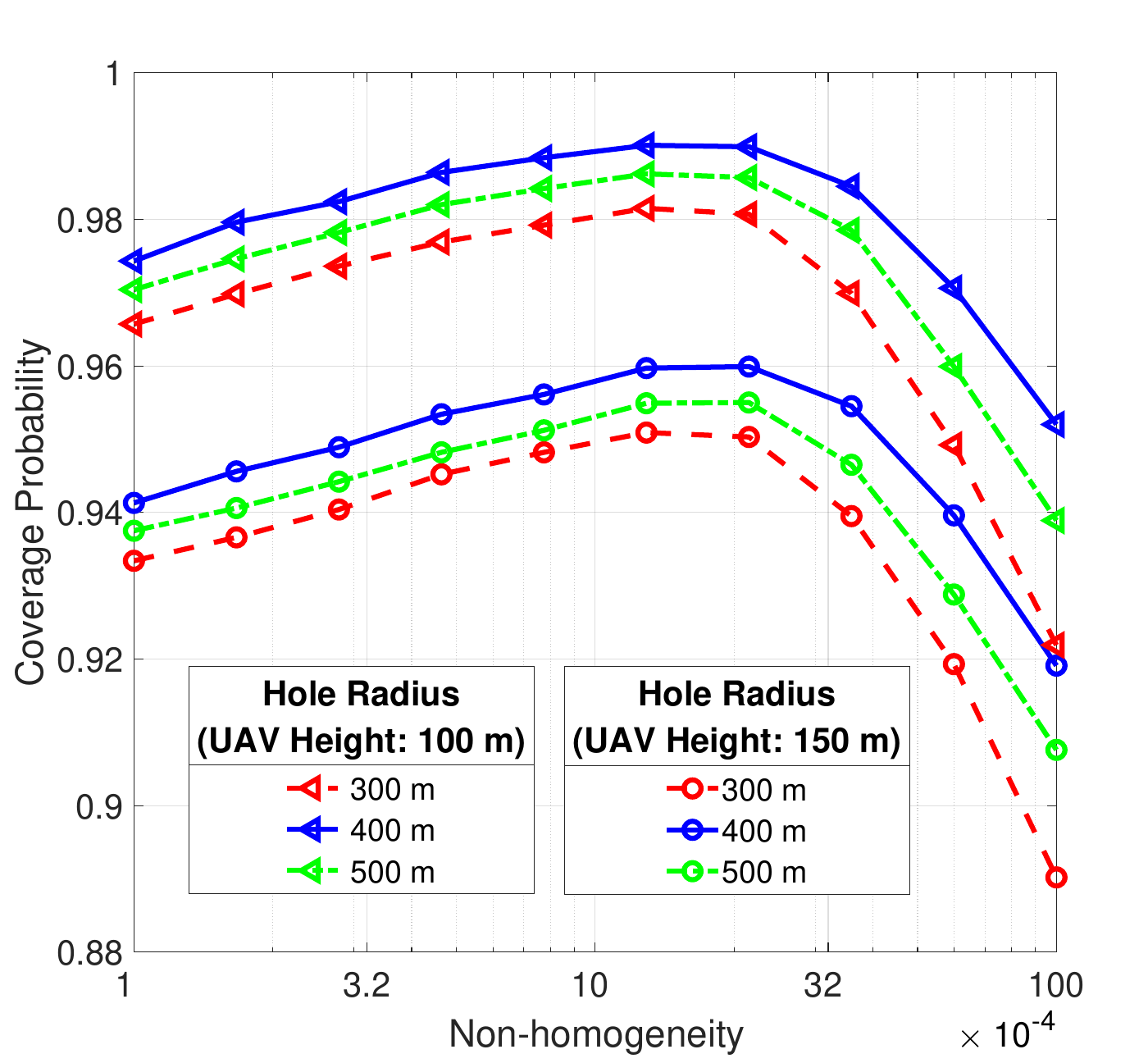}
\caption{Influence of different UAV distributions on coverage probability.}
\label{figure2}
\end{figure}

We assume that the number of UAVs follows a Poisson distribution with a mean of $14$. In this circumstance, besides height, there are two degrees of freedom that can be optimized in the UAV distribution. As shown in Fig.~\ref{figure2}, when the homogeneity and hole radius are known, the scaled density can be uniquely determined to maintain the mean number of UAVs at $14$. 

\begin{figure}[t]
\centering
\includegraphics[width=0.9\linewidth]{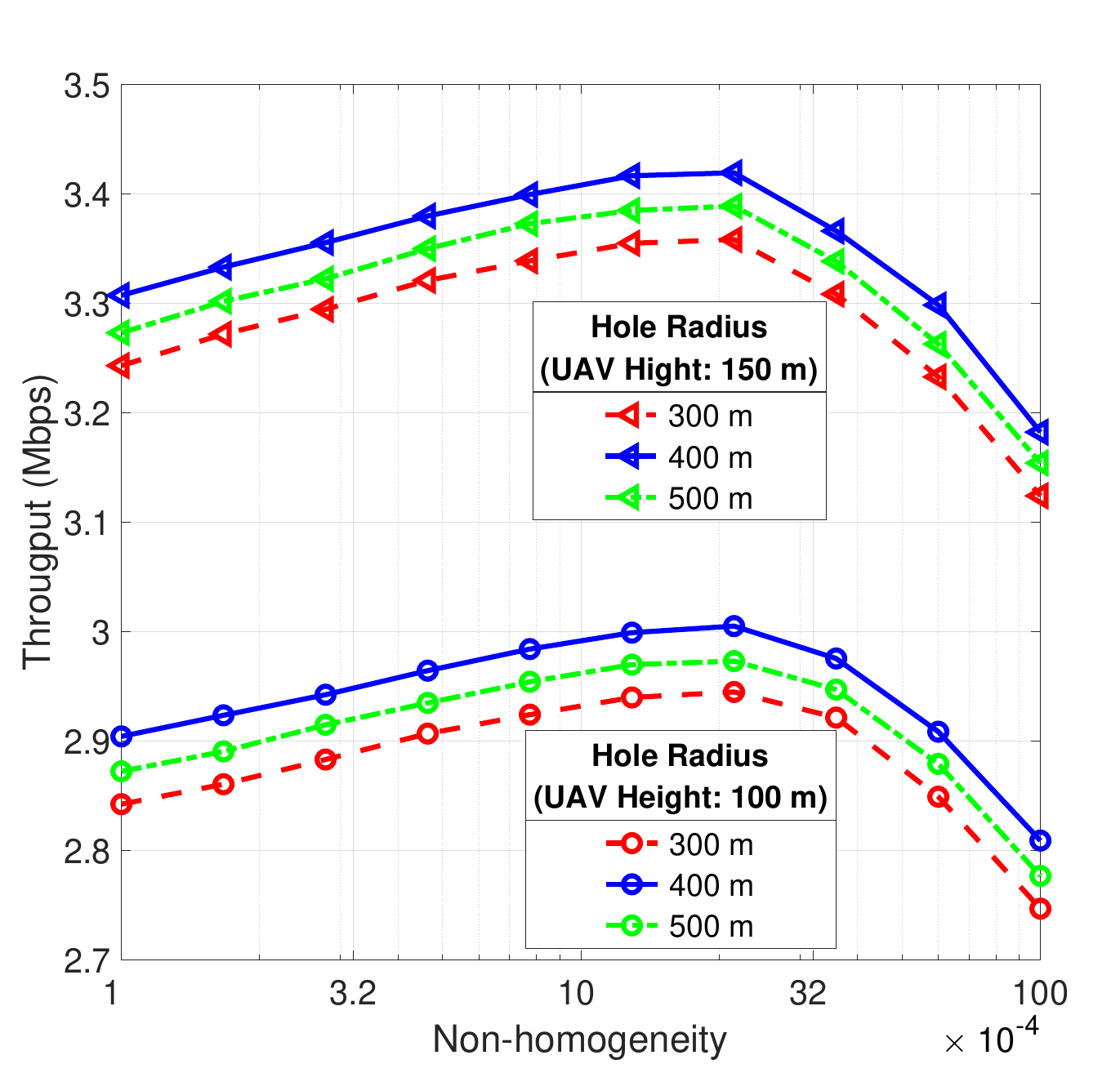}
\caption{Influence of different UAV distributions on throughput.}
\label{figure3}
\end{figure}

\par
Fig.~\ref{figure2} and Fig.~\ref{figure3} respectively investigate the impact of UAV distribution on coverage probability and throughput. Overall, the conclusions drawn from the two figures are similar. In comparison to the hole radius, the height of UAVs has a greater influence on communication metrics. With a fixed height, we can independently maximize coverage probability and throughput by optimizing the hole radius and non-homogeneity. In other words, while adjusting non-homogeneity to optimize communication metrics, the value of the hole radius has no effect on the optimal value of non-homogeneity. Furthermore, the hole radius and non-homogeneity pair that achieves globally optimal coverage probability and throughput are fundamentally consistent.

\subsection{Sensing Metric}\label{section4-3}
\begin{figure}[t]
\centering
\includegraphics[width=\linewidth]{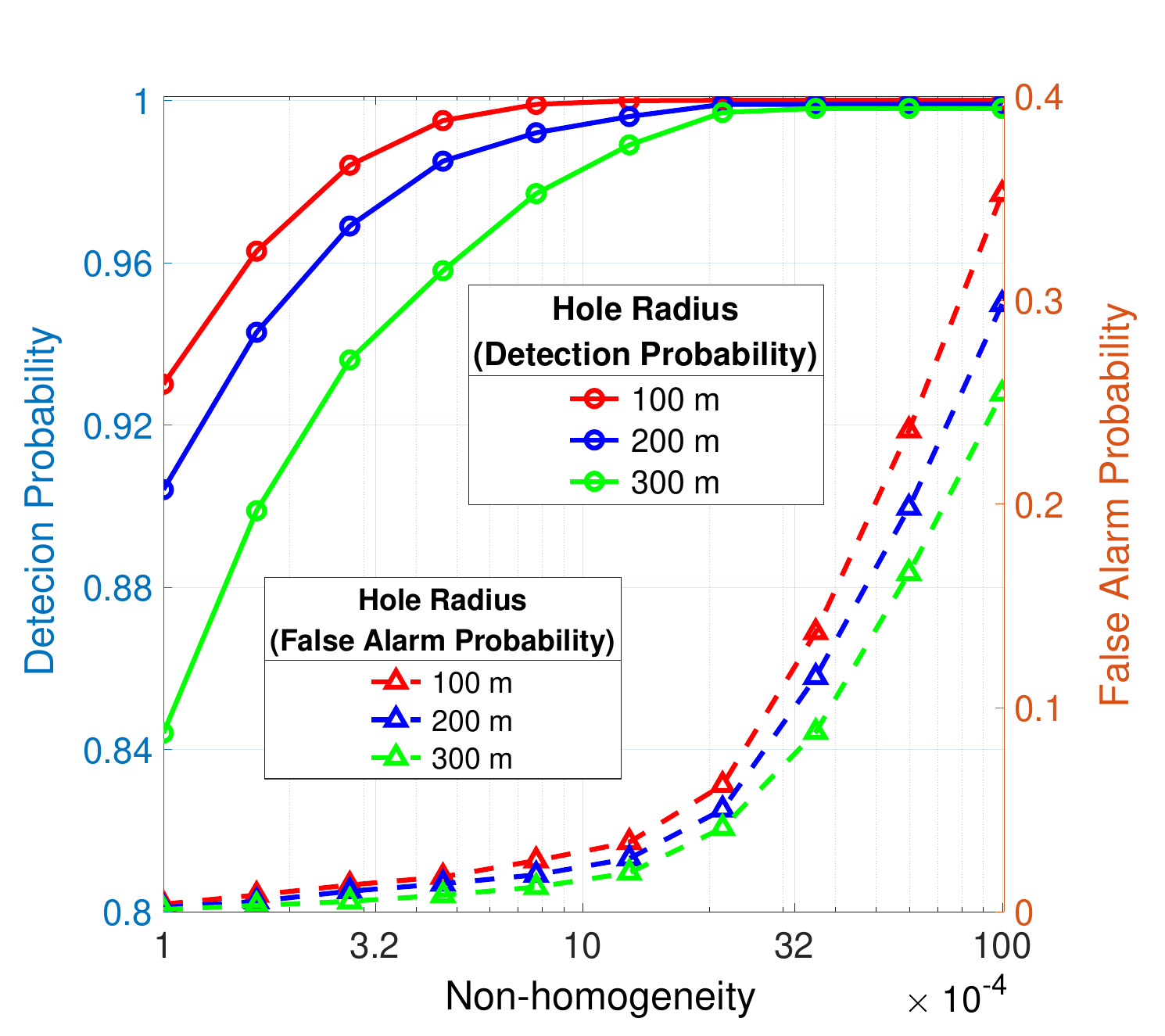}
\caption{Sensing metrics under different UAV distributions.}
\label{figure4}
\end{figure}

\begin{figure}[t]
\centering
\includegraphics[width=\linewidth]{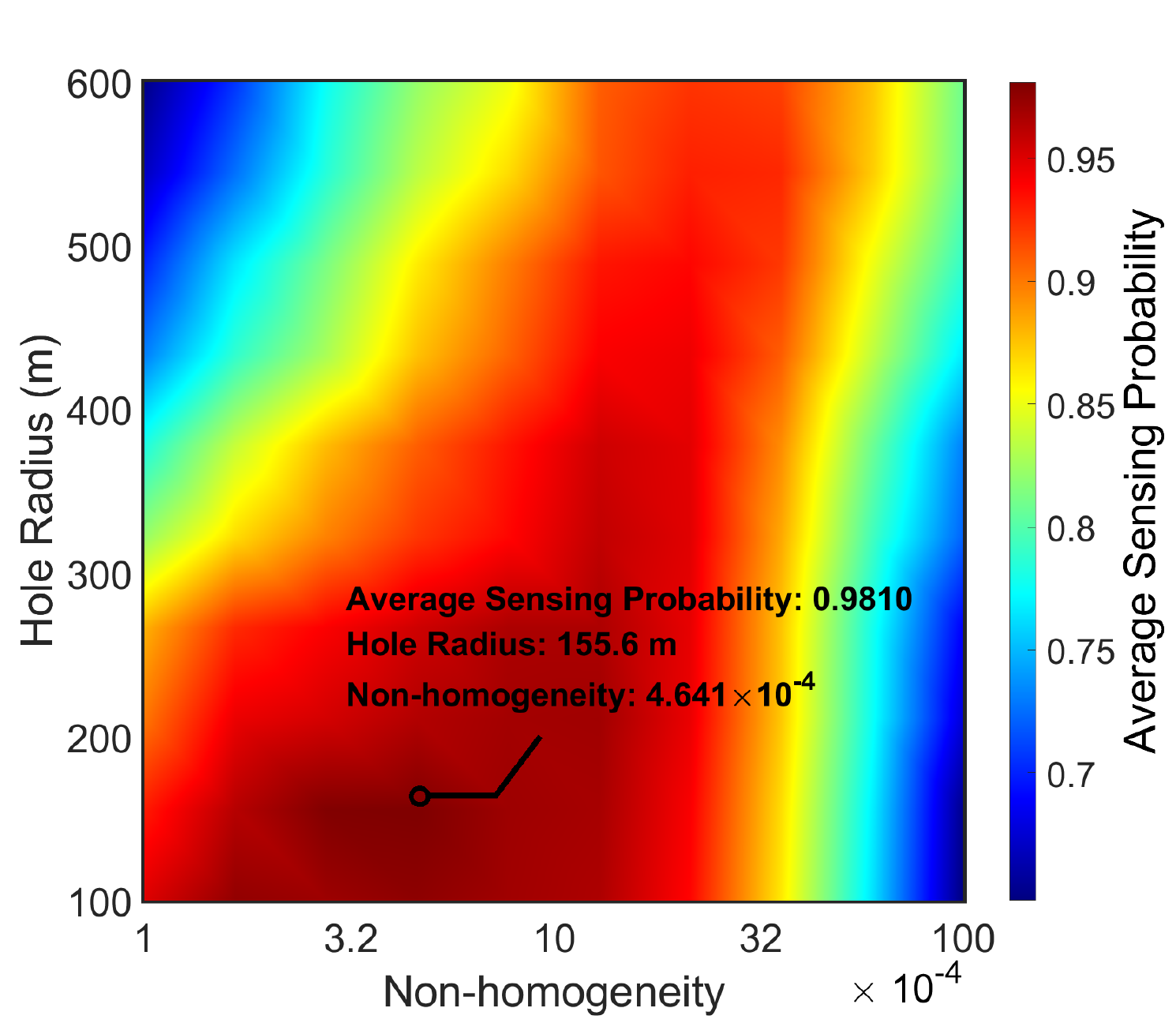}
\caption{Heat map of average sensing probability.}
\label{figure5}
\end{figure}

In Fig.~\ref{figure4} and Fig.~\ref{figure5}, the height of UAVs is fixed as $150$~m. As shown in Fig.~\ref{figure4}, increasing hole radius or decreasing non-homogeneity will lead to the overall UAV deployment in locations farther away from the BS, consequently reducing both the received power of target UAV and interference. Therefore, increasing the hole radius or decreasing non-homogeneity simultaneously decreases the detection probability and the false alarm probability. Furthermore, we can conclude that the optimal value for non-homogeneity is less than $4\times 10^{-3}$ because further increasing the non-homogeneity only increases the false alarm probability without detection performance improvement. Unfortunately, from the perspective of overall sensing performance, we can only observe qualitative conclusions from Fig.~\ref{figure4}. In other words, the optimal values of non-homogeneity and hole radius can not be identified.

\par
Therefore, we have generated a heat map, as shown in Fig.~\ref{figure5}, to further investigate the optimal UAV distribution with regard to sensing performance. We define the average sensing probability as a measure of overall sensing performance, mathematically defined as the product of detection probability and the complementary probability of the false alarm probability. The above results indicate that under a reasonable UAV deployment, the BS can almost guarantee detection while avoiding false alarms. 

\par
{  
This case analysis provides a solid example and meaningful insights for UAV scheduling. The maximum average sensing probability and its corresponding optimal UAV distribution parameters are marked in Fig.~\ref{figure5}. We can implement the initial deployment of the UAV network based on this set of optimal parameters. Subsequently, during the actual network operation, further interference mitigation can be achieved through fine-grained scheduling, such as maintaining UAV spacing. Using a similar heatmap visualization approach, we can also test network performance and facilitate cell planning by adjusting the maximum radius of UAV distribution.}

\section{Conclusion and Open Issues}\label{section5}
We presented a comprehensive SG-based framework for analyzing the performance of ISAC networks at the network level, which consists of modeling and analysis. From the modeling perspective, we explained how to employ distribution models and channel models in ISAC network using SG tools. From the analytical perspective, we described the categories of network types, performance metrics, and research objectives. With regard to aspects that have not been thoroughly investigated, we propose a case study to further explore the advantages of SG in analyzing ISAC networks. The hybrid network formed by a BS and several UAVs provides network coverage for residents, with BS controlling the deployment of UAVs through its sensing capabilities. By accurately simulating the distribution of residents and BS as well as the urban channel environment, we optimize the communication and sensing performance of ISAC networks by adjusting the deployment of UAVs. 

\par
{ 
As the integration of ISAC and SG is still in the exploratory phase, scholars are focusing on establishing the most fundamental analytical framework. If accuracy in spatial distribution modeling and channel modeling can be improved, it is believed that the accuracy of performance analysis can be further enhanced. The second direction for future development is exploring more joint metrics. As mentioned, ISAC networks have significant advantages in spectrum efficiency, energy costs, and hardware costs. However, currently only PSE has been analyzed, and there are currently no suitable parameters for measuring price-related aspects within the SG framework. Finally, more typical applications and network types are worthy of research, such as satellite networks. Satellite networks have the demand for applying ISAC technology, and SG has been confirmed to enable simplified but accurate constellation modeling, as well as accurate performance analysis.
}

\bibliographystyle{IEEEtran}
\bibliography{references}

\end{document}